# Collapse and cavitation during drying of water-saturated PDMS sponges with closed porosity

Phu Tuan Anh Nguyen[1,2], Matthieu Vandamme[1], Artem Kovalenko[2*]

[1] – Navier, Ecole des Ponts, Univ. Gustave Eiffel, CNRS, Marne-la-Vallée, France

[2] – Laboratoire Sciences et Ingénierie de la Matière Molle, ESPCI Paris, PSL University, Sorbonne Université, CNRS, F-75005 Paris, France

* - corresponding author

**Abstract:**
In this paper, we study the drying of water-saturated porous polydimethylsiloxane (PDMS) elastomers with closed porosity in which the evaporation of water is possible only via the diffusion across the PDMS. Starting from water/PDMS emulsions, we fabricate soft macroporous samples with different pore diameter distributions and average diameters ranging from 10 to 300 µm. In these materials, the drying may lead either to a collapsed state with low porosity or to the cavitation and reopening of a fraction of the pores. Using optical microscopy and porosity measurements, we showed the importance of the pore diameters and interactions on the result of drying. At pore diameters lower than 30 µm, the majority of pores remain collapsed. We attribute the permanence of the collapse of most small pores to a low probability of cavitation and to the adhesion of the pore walls. Pores with diameters larger than 100 µm tend to reopen after the water they contain cavitates. The behavior of pores with diameters ranging from 30 to 100 µm depends on the porosity and drying temperature. We also visualize collective cavitation upon drying of sponges initially saturated with a sodium chloride solution. In this case, the cavitation in the largest pores leads to reopening of small pores in a neighboring zone of the sample. To our knowledge, our results present the first experimental proof of the pore-size-dependent and cooperative nature of the response of soft sponges with closed porosity to drying.

# 1. Introduction

Porous polymers are widely used for acoustic and thermal insulation. In recent years, increasing attention is paid to soft porous elastomers, or sponges[1], due to their use as compressible parts in mechanical sensors and soft actuators. Moreover, these materials were used to fabricate novel acoustic metamaterials[2] and gradient lenses[3] due to extremely low values of the sound speed obtained at ultrasound frequencies[4]. To achieve a controlled porous structure, a template consisting of liquid droplets[5] or solid particles[6] is dispersed in the liquid prepolymer, usually polydimethylsiloxane (PDMS), which is then crosslinked to obtain an elastomer. The template is then removed by etching[7] (in the case of solid particles), washing and drying to obtain pores with sizes and volume fraction similar to that of the template. Ideally, the drying step should lead to air-filled porous materials with low mechanical moduli. However, it was reported previously that capillary forces during drying may lead to a collapse of the porosity and to a loss of the desired acoustic or mechanical properties[8,9]. One solution to avoid this effect is to generate open porosity by using adhesive emulsions with aggregated droplet structure[9]. The micron-sized "windows" between pores allow water evaporation and air invasion in the pores. This method suffers of the sensitivity to the emulsion fabrication conditions and spatial inhomogeneity of the aggregated porous structure. The second strategy is to use supercritical drying[3,8]. This approach requires consecutive washing of the samples with an intermediate solvent (for example, ethanol) and liquid carbon dioxide which is then removed by supercritical cycle at high pressure. While it allows fabrication of materials with closed porosity and microscopically homogeneous pore distribution, such process is expensive and time-consuming. Thus, it would be interesting to propose alternative approaches of drying for soft sponges with closed, homogeneously distributed porosity, which is impossible without rationalization of the collapse problem. However, neither the collapse mechanism during thermal drying nor the dependence of the final porosity on material parameters and drying conditions are yet well understood.

Interestingly, Milner et al.[10] has recently shown that an "isolated" water-filled pore in PDMS with the diameter $d \sim 1$ mm reopens during drying via cavitation (i.e., via formation of a bubble inside the water it contains). The authors observed the shape of the pore and showed the presence of several drying stages. In the first so-called "breathing" mode, the pore contracted homogeneously and remained spherical. Next, a creasing instability was observed at the pore surface which leaded to a non-spherical pore shape. According to Cai et al.[11], this instability is preferred to buckling for thick pore walls (corresponding to low porosities for sponges). The number of creases depends on the surface energy of the pore and on the high-strain mechanical properties of the elastomer. The next drying stage which takes place at a certain moment before or after creasing is cavitation of the water it contains[10,12] (i.e. nucleation of a water vapor bubble). This stage is followed by the release of the (negative) pressure inside the pore and by an expansion of the pore back to its initial size and shape. By tackling the bubble expansion velocity during this last stage, Bruning et al. estimated[12] the cavitation pressure $p_{cav} \approx -1.4$ MPa, which absolute value is much lower than 20-30 MPa found for water in rigid monodisperse pores by Vincent et al[13] and synthetic trees by Wheeler et al[14]. This difference was explained by the highly hydrophobic nature of the PDMS which favors heterogeneous bubble nucleation.

Cavitation is the necessary but not sufficient condition for pore reopening. The stress in the matrix should be high enough to promote the cavity growth. This effect was intensively investigated in the literature related to such cavity growth in adhesives and elastomers[15], which is also called "cavitation" although no liquid is involved. We will refer to this process as to "solid cavitation", to avoid any ambiguity with the cavitation of the pore water in water-saturated porous solids, which was described in the previous paragraph. It was shown that a small pore inside an elastic matrix with Young's modulus $E$ is unstable and will grow to the infinite size at mean tensile stresses higher than $\frac{5}{6}E$: such dependence of the stress at which "solid cavitation" occurs on the modulus was observed experimentally[15–17]. For the analysis of the stability of small pores, the polymer surface energy should be taken in account. Gent and Tompkins[16] introduced the surface energy term in the mechanical model of the pore deformation. They showed that the critical mean stress for the infinite inflation of the pore increases with a decreasing initial radius of this pore and attains hundreds of MPa for radius of few nanometers. The characteristic length scale of the action of surface forces is called the elasto-adhesive length $\gamma/E$, where γ is the surface energy of the elastomer. For a typical PDMS elastomer with γ = 40 mN/m and E ~ 1 MPa, the elasto-adhesive length is 0.04 μm. The model of Gent and Tomkins shows that a significant impact of surface energy on the tensile stress at which pore growth occurs is observed for pore dimensions lower than 0.1 μm[16,17]. For larger cavities, the growth should be observed at stresses of the order of the modulus $E$, i.e. about 1 MPa. This explains why the individual large pores in PDMS observed by Milner et al[10] and Bruning et al[12] reopen after cavitation at $p_{cav} \approx$ -1.4 MPa.

The aim of this study is to characterize experimentally the effect of the size and number of pores on how porous elastomers with closed porosity initially saturated with water or a salt solution behave during drying. Using previously reported emulsion-templated protocols[4], we synthesize macroporous polydimethylsiloxanes (PDMS) with pore diameters ranging from 1 to 100 μm. First, we use optical microscopy to observe the drying of "diluted" pores (5% of porosity) in thin (i.e. with a 0.5 mm thickness) water-saturated samples. We show that pore behavior differs regarding their diameters: after their total shrinkage, pores with diameters larger than 30 μm are most likely to reopen by cavitation while the major part of smaller pores remain in a collapsed state for at least 16 hours of observation. The part of reopened pores increases with increasing the drying temperature from 60°C to 110°C. Next, by using density measurements to measure porosity, we compare the drying of thick (i.e., with a thickness of about 10 mm) water-saturated samples with 30% porosity and two different pore size distributions. After drying, the sample with small pores ($d$ ~ 70 μm) is almost completely collapsed while the sample with larger pores ($d$ ~ 300 μm) shows reopening of a significant fraction of its pores. Finally, we show the cooperative nature of the cavitation of pores in samples initially containing sodium chloride solution in the pores. We observe that the opening of larger pores leads to reopening of neighboring small pores. To our knowledge, the observed size-dependent and cooperative effects were not reported before and present an interest for the soft matter physics.

## 2. Experimental

*2.1. Materials*

A two-component Sylgard®184 PDMS kit was supplied by Neyco. It is composed of vinyl-modified PDMS base (SiVi) and silane-modified PDMS curing agent (SiH). We used deionized MiliQ water and NaCl from Sigma Aldrich.

*2.2. Synthesis of porous PDMS samples*

We used the previously reported emulsion-templated method of synthesis[4]. The continuous phase of the emulsions was obtained by mixing SiVi and SiH parts in the 10:1 ratio. After mixing, the continuous phase was degassed at room temperature to remove bubbles. The water phase was either MiliQ water or 1.5 wt.% NaCl solution. Surprisingly, stable emulsions with 5% or 30% volume fraction of water could be produced without adding surfactant. We explain this by the high viscosity of the PDMS and the presence of low-molecular species[18] that can act as surfactants. The corresponding amount of water was incorporated in about 40 g of PDMS under gentle mixing with a spatula (at approximately 1 tour per second). The mixture was next stirred for 2 minutes via a mechanical mixer Heidolph RZR 2162 with a helical geometry with diameter of 3 cm. The stirring conditions are given in Table 1.

By controlling the stirring speed of the emulsions and, consequently, the shearing stress, one should be able to tune the size of the droplets. However, because of the high viscosity of the Sylgard 184, it was impossible to obtain narrow size distributions using simple mixing devices. After several tests, we determined empirically stirring protocols giving samples with significantly different drying properties. The pore diameter distributions $N_i(d_i)$ were obtained by manual pore tracking of optical microscopy images on at least 1000 pores using ImageJ software. The pore distribution by volume $\varphi_i(d_i)$ was calculated using the equation:

$$\varphi_i(d_i) = \frac{N_i d_i^3}{\sum_i N_i d_i^3} \qquad (1)$$

where $N_i$ are the numbers of pores with the diameters $d_i$.

To obtain samples for microscopy, the volume fraction of the aqueous phase was fixed to 5%. The emulsion was then poured between two PET-coated glass molds with a 0.5 mm rubber spacer and the molds were introduced in a sealed metallic box and heated at 60°C for 14 hours. These samples are designed as "thin" in the rest of the paper.

For density measurements, we fabricated samples starting from a 30% emulsion. The emulsion was poured in 60 mL plastic tubes and centrifuged at 1000 rpm for 5 min to remove gas bubbles. After polymerization at 60°C for 14 hours, the samples were unmolded and cut into equivalent pieces with an average weight of 5 g and thickness of about 10 mm. These samples are designed as "thick" in the rest of the paper.

In Table 1 we summarize the composition, stirring conditions and mean pore diameters for the samples and the methods we used to study their drying.

*Table 1 – Characteristics and drying investigation methods of the fabricated sponges*

| Sample name | Initial porosity | Stirring speed | Volume-weighted mean pore diameter | Aqueous phase | Optical microscopy (thin samples) | Porosity measurements (thick samples) |
|---|---|---|---|---|---|---|
| D200-5 | 5% | 375 rpm | 200 μm | MiliQ water | + | - |
| D10-5 | 5% | 1500 rpm | 10 μm | MiliQ water | + | - |
| D300-5NaCl | 5% | 375 rpm | 300 μm | 1.5% NaCl | + | - |
| D50-5NaCl | 5% | 800 rpm | 50 μm | 3.65% NaCl | + | - |
| D300-30 | 30% | 200 rpm | 300 μm | MiliQ water | - | + |
| D70-30 | 30% | 375 rpm | 70 μm | MiliQ water | - | + |

*2.3. Observation of drying of thin samples with 5% porosity*

The samples with 5% initial porosity and thickness of 0.5 mm were placed in a heating microscope stage HFS 91 (Linkam) under an optical microscope. The cell was connected to an air pump which maintained a flow of dry air equal to 0.5 liters/min through the chamber. After rapid heating to the desired temperature (about 1 min), the recording of images was performed every 10-30 s using a ×10 objective. Two drying temperatures, 60°C and 110°C, were used for each sample. The drying of the sample D200-5 was also performed at 90°C.

*2.4. Drying and characterization of thick samples with 30% porosity*

For thick samples (about 10 mm thickness), we desired to perform the drying at a well-controlled temperature and without any solid support which could create a tangential stress affecting the sample's shrinkage. Thus, the drying was done in a hot glycerol which homogenizes rapidly the temperature and does not diffuse into the pores (and hence does not swell the matrix). A 50 mL vial was filled with dry glycerol and heated to the desired drying temperature. Next, the sample was immerged in the bath for 7 days. This time was enough for complete drying, as deduced from the fact that the relative weight loss reached an asymptotic value close to the initial water mass fraction, with the difference between the relative weight losses at the 7$^{th}$ and 8$^{th}$ days being less than 0.2 %. The initial porosity of the samples was calculated from the amount of water lost during drying. Similar values, very close to 30%, were obtained for specimens extracted from the top and from the bottom of the centrifugation tube. Thus, we concluded that centrifugation did not create measurable gradients in the volume fraction of the emulsion.

The density of dried samples was measured by using a home-made hydrostatic balance which measures the force necessary to immerge a sample in a liquid using a thin rigid wire. To increase the precision, we used several liquids including pure water and NaCl solutions with concentration ranging from 3 to 7 wt.%. The density was calculated using the balance of the forces acting on the immerged sample:

$$m_{sample} + \Delta m_{pull} = \rho_{liquid} V_{sample} \quad (2)$$

We thus infer:

$$\frac{\Delta m_{pull}}{m_{sample}} = \frac{\rho_{liquid}}{\rho_{sample}} - 1 \quad (3)$$

where $m_{sample}$, $V_{sample}$ and $\rho_{sample} = m_{sample} / V_{sample}$ are the mass, the volume, and the mass density of the sample, respectively; $\rho_{liquid}$ is the mass density of the liquid and $\Delta m_{pull}$ is the difference between the mass measured when the sample is forcibly immersed in the liquid by plunging the thin rigid wire and the mass measured in absence of sample but when the thin rigid wire is plunged in an identical position. The sample density was obtained from an affine fit of the curve of $\Delta m_{pull}/m_{sample}$ as a function of $\rho_{liquid}$, since, as Eq. (3) shows, the slope of this fit must be equal to $1/\rho_{liquid}$ (the curves are given in the SI-1a). By using the same method, we measured the density of the PDMS matrix $\rho_{PDMS}$ = 1034 kg/m$^3$, which is slightly lower than the reported value[18] of 1100 kg/m$^3$. To calculate the sample porosity $\Phi = V_{pores}/V_{sample}$ (where $V_{pores}$ is the volume of pores in the sample), we used the following equation:

$$\Phi = 100\% * \left(1 - \frac{\rho_{sample}}{\rho_{PDMS}}\right) \quad (4)$$

*2.5. Other characterizations*

The microstructures of the obtained porous samples were characterized with a TM-1000 scanning electron microscope (Hitachi). Prior to imaging, the samples were fractured in liquid nitrogen, dried at 60°C and covered with a thin gold layer by vapor deposition. The observation was performed on the fractured surface.

Tensile loading measurements on parallelepipedal samples of about 30×5×0.5 mm were performed using an Instron 5565 apparatus. In the linear elastic regime, the dependence of the axial force $F$ with the axial strain $\varepsilon$ was obtained and treated according to Hooke's law:

$$\sigma = E\varepsilon \quad (5)$$

where $\sigma = F/S$ is the axial stress, $S$ the initial cross-section area of the sample, and $\varepsilon = (L-L_0)/L_0$ with $L_0$ and $L$ the sample's length before and during the test, respectively.

## 3. Results and discussion

*3.1. Structure and mechanical properties of the PDMS sponges*

For both the synthesized samples with 5% and 30% of the dispersed water (or salt solution) phase, the porosity is closed. This is confirmed by the electronic microscopy images of the sample sections shown on Figure 1a and b. The fact that the pores are less spherical than on the optical microscopy images (see next sections) may be due to the cutting and drying-induced stresses. The closed porosity is formed because of the isolated droplet structure of the emulsion[9]; moreover, at 5% and 30% of porosity, the pore walls are rather thick and are not connected by macroscopic windows between them. Like in the case of a unique water-filled pore[10], the drying of such samples happens by pervaporation through PDMS and not by the invasion of the pore network by a liquid-vapor meniscus.

We also checked that macroscopic mechanical properties of water-filled porous samples are qualitatively similar to the non-porous PDMS, as confirmed by the stress-strain curves on Figure 1c, which exhibits curves with similar shape. The Young's modulus of the wet non-porous PDMS is about 1.6 MPa and decreases to 1.16 MPa and 0.98 MPa for water-saturated porous samples with porosities equal to 5% and 30 %, respectively. The decrease of modulus for the water-saturated porous samples with respect to the dry bulk PDMS may be explained by the presence of pores and by the influence of humidity on polymerization. The fact that the difference in modulus between the wet non-porous PDMS and any of the two water-saturated porous samples is larger than that between the two water-saturated porous samples suggests that, for our samples, the modulus is more significantly impacted by humidity than by the presence of pores. Water may hydrolyze some of the Si-H groups in the curing agent and decrease the degree of curing[19]. However, the effect is minor and should not drastically impact the interpretation of the results in next sections.

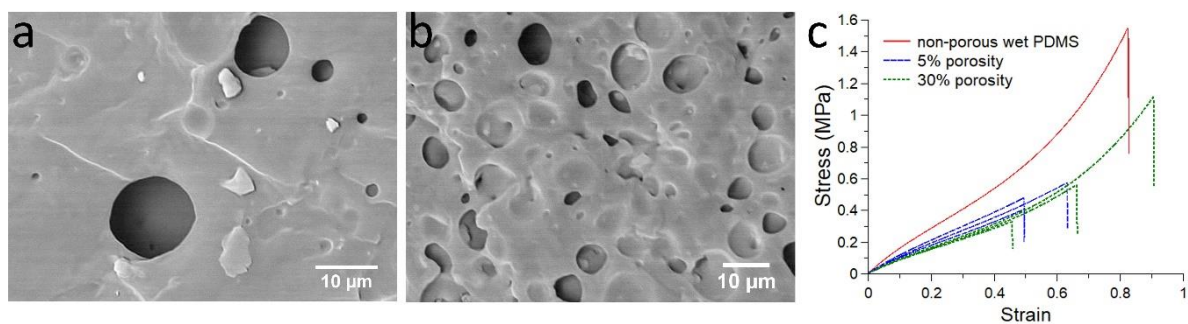

*Figure 1 – a) and b) Scanning electron microscopy images of sections of the sample D200-5 with 5% porosity (a) and sample D70-30 with 30% porosity (b). c) Stress-strain curves of the same samples in wet state in comparison with a non-porous PDMS "swollen" in water at 60°C for 6 hours (red curve). For each type of porous sample, the measurement was performed on 3 samples.*

*3.2. Visualization of the pore collapse and reopening*

First, we performed a microscopic observation of drying for the sample D200-5 with 5% porosity and very polydisperse pore diameter distribution (red bars on Figure 2a and b). The corresponding optical microscopy image in the initial water-saturated state is given on Figure 2c. When counting by number, the distribution is centered at about 20 μm, however, the volume-weighted distribution obtained via eq 1 is centered at about 200 μm. The volume-weighted distribution is more relevant for the porosity Φ which is, by definition, the volume fraction of pores.

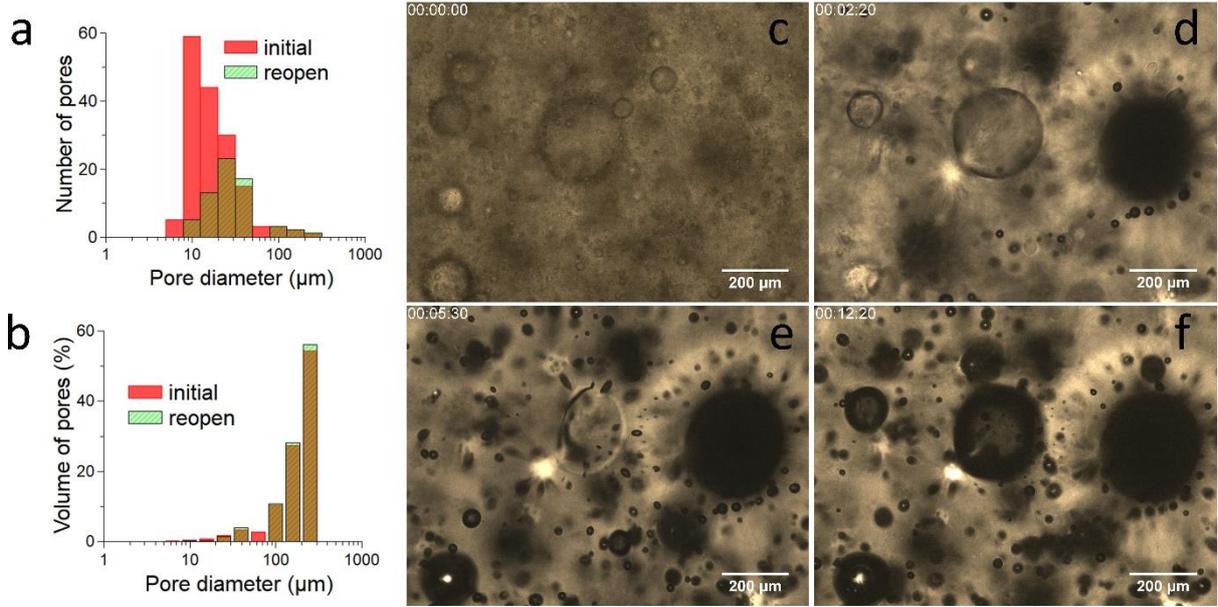

*Figure 2 – a) and b) Distributions of pore diameters in the D200-5 sample obtained for initial, water-filled pores (red) and pores reopened via cavitation (semi-transparent green); c-f) optical microscopy images of the D200-5 sample during drying at 110°C. The overlap of the pore diameter distributions between water-filled and reopened pores appears as brown.*

The images on Figure 2d-f show the microscopy images of pores in the D200-5 sample when dried at 110°C. For a video-stack of drying images, see the SI-2. First, we observe a rapid shrinkage of small pores which seem to disappear (Figure 2d). The pores with larger diameters also decrease in volume and show a transition to a creased state, as observed for the central pore on the Figure 2e. We define the time of collapse as the moment when pores are non-detectable. For larger pores the dissolution time is longer. This may be explained with the diffusive model of the pore shrinkage proposed by Milner et al.[10] and Bruning et al[12]. In this model, the diameter of the pore scales with time $t$ as $d = d_0\sqrt{1 - kt/d_0^2}$ where $d_0$ is the initial diameter and the kinetic factor $k$ is mainly governed by the diffusion coefficient and the difference between the water concentration $c_{eq}$ in PDMS near the pore and $c_\infty$ near the sample's edge. In absence of impurities (such as salt), the values of $c_{eq}$ and the diffusion coefficient depend only on temperature. The concentration $c_\infty$ on the sample's edge is determined by the relative humidity in the chamber. Hence, we may suppose the value of $k$ to be similar for all pores. The dissolution time scales as $t_{collapse} \sim d_0^2$, which explains the fact that the small pores reach full shrinkage faster than the large pores. However, quantitative analysis of the porous sample would not be straightforward, because the model was developed for a single pore and does not include interaction between pores with different drying dynamics.

Some pores reopen via cavitation, giving pores of diameters similar to the initial ones and located at places similar to the initial ones (compare the final state on Figure 2f and the initial sate on Figure 2c). The reopened pores show a higher contrast than in the water-saturated state because of the difference of the optical index between PDMS and water vapor. On Figure 2a, the comparison between the diameter distribution of the reopened pores and that of the pores before drying shows that most pores with diameters lower than 30 μm remain collapsed while the majority of larger pores reopen. However, on the volume-weighted diameter distribution (Figure 2b), the contribution of the small pores is almost invisible, meaning that their collapse will not strongly affect the total pore volume and hence the porosity. The fact that the number of reopened pores with a diameter around 40-50 µm is higher than the number of pores observed before drying may be caused by the change of the transparency of the sample during drying, because of which deeper pores may become visible. Because of the small pore size, the optical microscopy doesn't allow to determine precisely the nature of the cavitation (liquid or "solid" one). The results of Bruning et al[12] on millimeter-scale pores in PDMS demonstrate that the cavitation occurs in the water phase and the pore reopening happens about 0.1 ms after cavitation. We did not observe any correlation between different cavitation events: reopening happened some time after the complete shrinkage of a given pore, with no obvious trend.

For a second experiment, we chose the sample D10-5 which contains exclusively the small pores (Figure 3a) with an average diameter of about 10 µm (the diameter distribution was obtained from the image of a thin layer of the emulsion used for the synthesis given in the SI-1b). In this case, the imaging of pores is very challenging because of their small diameters and the change of the transmittance of the sample with time (see the video-stack of drying images in the SI-3). However, once the pores reopen, their detection is quite easy. Only pores that were located near the focus plane and that appeared as black spots were counted as reopened. First, we performed a drying at 60°C, far below the boiling point of water. The images show that most pores remain collapsed and invisible in the shrunk state while only few pores reopen (Figure 3b). In the present configuration of the microscope, the effective depth of field (DOF), i.e. the distance which allows simultaneous detection of the reopened pores, was roughly estimated by changing the focus distance as DOF ~ 200 µm. Taking the image with lateral dimensions H×W = 1200×900 µm$^2$, we assumed that pores are spherical and calculated the volume of the reopened pores $V_{pores}^{60°C} \approx 3*10^5 \mu m^3$. The estimated porosity is therefore as $\Phi^{60°C} \approx V_{pores}^{60°C} / (DOF*H*W) \approx 0.15$ %. This value is at least one order of magnitude lower than the initial porosity equal to 5%. This confirms that the majority of the pores collapsed and did not reopen. The number of reopened pores does not change after 30 min of drying for at least 16 hours (Figure 3c). The background on the image at 16 hours appears darker and less homogeneous because of the low light intensity used to acquire this image. We also give images of initial and dry states at higher magnification in SI-1c.

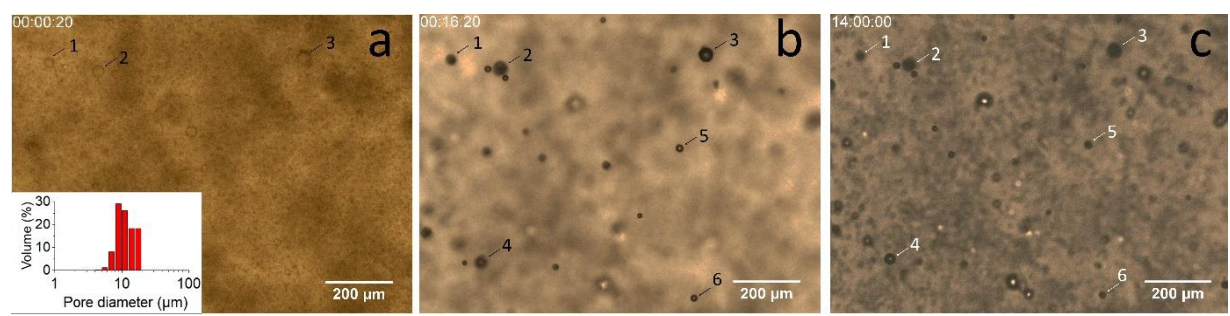

*Figure 3 – Optical microscopy images of the D10-5 sample during drying at 60°C: a) initial state (small pores are poorly visible); b) state after 16 min of drying; c) state after 16 hours of drying. The insert on the image a) shows the pore diameter distribution in the initial state. Some clearly identified pores are labeled with numbers.*

The D10-5 sample dried for 16 hours at 60°C was next heated to 110°C, which is above the boiling temperature of water. Figure 4 and SI-4 show that heating leaded to reopening of another small fraction of pores. Following the same methodology as at 60°C, the final porosity at 110°C may be estimated at about 0.3%, which is well below the initial one. This analysis confirms that small pores remain shrunk at least until the end of the experiment.

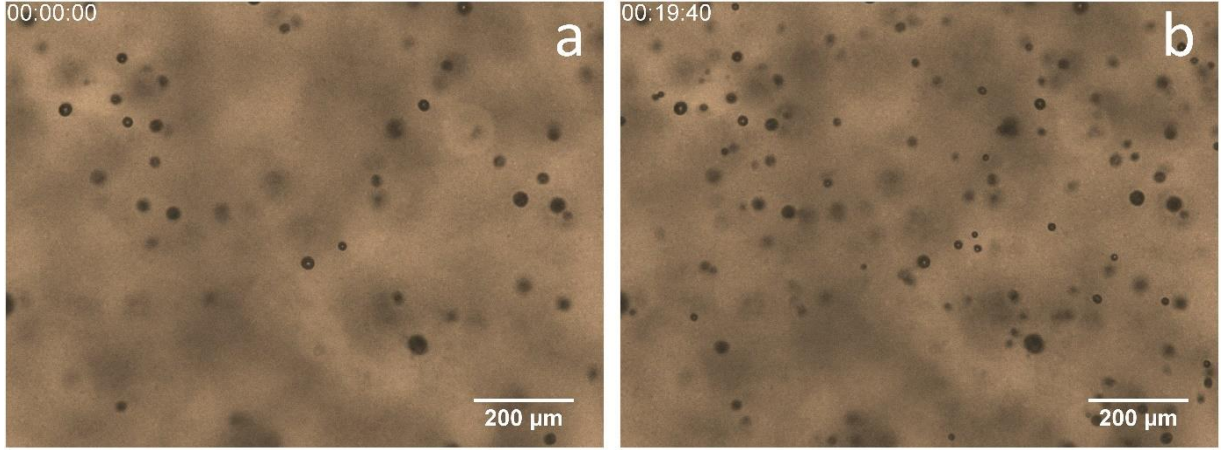

*Figure 4 – Optical microscopy images of the same area in the sample D10-5 before (a) and after (b) increasing of the drying temperature from 60°C to 110°C.*

*3.3. Macroscopic porosity measurements*

We compare the samples D70-30 and D300-30 which have similar initial porosity $\Phi_{init} \approx 30\%$ and different pore sizes. Figure 5 shows the optical images of a cross-section of the samples. The number-weighted distributions as well as some additional images are given in the SI-1d. In contrast to the transparent samples with 5% of porosity (section 3.2), the samples are opaque and the observation of the pores is possible only near the cross-section plane. The probability of a pore to be found in the cross-section is inversely proportional to its diameter[20] and thus $N_i \propto N_i^{section}/d_i$, where $N_i^{section}$ are the numbers of pore of diameter $d_i$ observed in the cross-section and $N_i$ the number of the pores of diameter $d_i$ in the sample. Consequently, the volume fraction $\varphi_i$ of pores of diameter $d_i$ is given by:

$$\varphi_i(d_i) = \frac{N_i d_i^3}{\sum_i N_i d_i^3} = \frac{N_i^{section} d_i^2}{\sum_i N_i^{section} d_i^2} \qquad (6)$$

. The obtained distributions are displayed in Figure 5c.

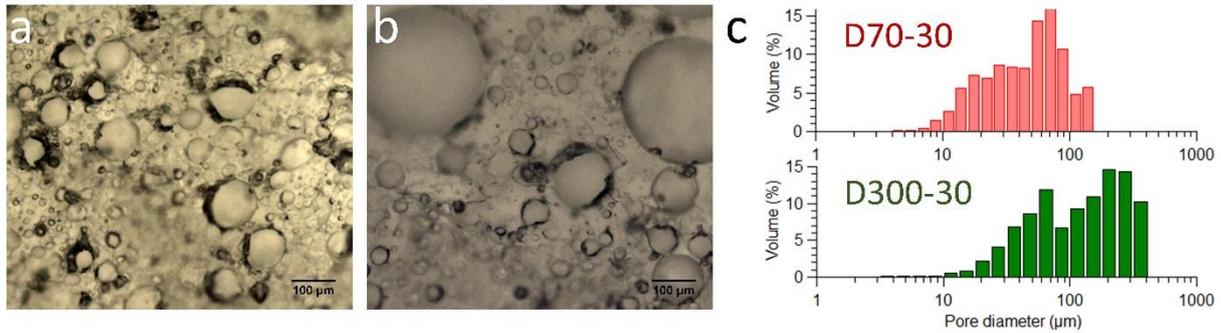

*Figure 5 – Optical microscopy images of a cross-section of D70-30 (a) and D300-30 (b) samples and the corresponding volume-weighted diameter distributions (c).*

The data of the samples dried in a glycerol bath at 60°C or 110°C are presented on Figure 6a. For both temperatures, the drying loss (which is equal to the ratio between the total volume of evaporated water and the initial volume of the sample, and is therefore an estimate of the initial porosity) was about 30%. Next, we show the porosity values calculated from density data obtained using the hydrostatic balance. The D70-30 sample's porosity after drying is very low, slightly increasing with drying temperature from 0.5 to 3%. By contrast, the D300-30 sample shows the reopening of almost a half of the pores (by volume). Again, the porosity increases with the drying temperature. The difference between the two samples is demonstrated by the flotation test in water on Figure 6b: the porous D300-30 sample rapidly rises to the surface while the collapsed D70-30 sample sinks. Note that we did not detect any change in the porosity after several months of storage at room temperature.

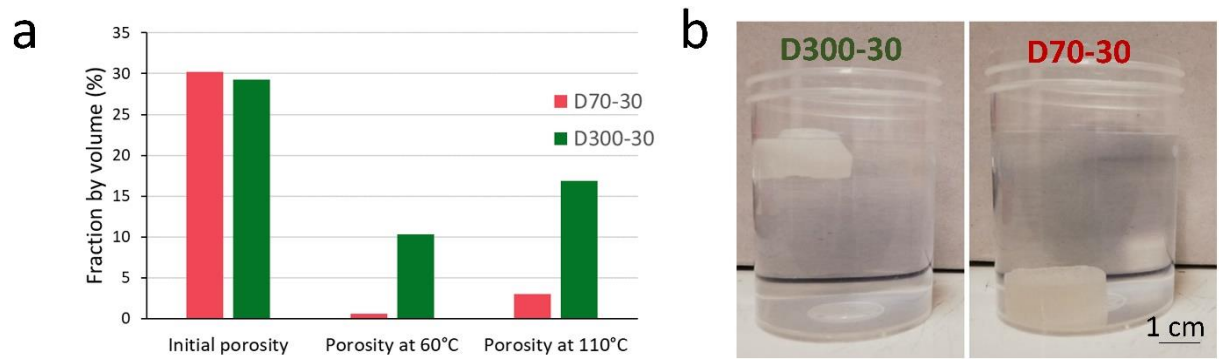

*Figure 6 – a) Characteristics of the dried thick samples with different pore sizes; b) a photograph of the dried samples demonstrating flotation (D300-30) and sinking (D70-30) in pure water.*

*3.4. Factors affecting probability of reopening*

The results of microscopy (Section 3.2) and macroscopic density measurements (Section 3.3) confirm that the pores with larger diameters are more likely to cavitate and reopen than small pores. Because of the large pore size distribution, it is difficult to demonstrate the existence of a critical pore diameter above which pores would reopen and below which pores would not. Also, we may expect the cavitation to be probabilistic and to depend not only on the pore diameter but also on small deviations of the pore shape and on the proximity to the neighboring pores.

We repeated the drying several times on different pieces of the sample D200-5 at temperatures of 110°C and 90°C to obtain a statistically relevant measurement of the fraction of pores of a given diameter that reopen during drying. The results at 110°C are displayed in Figure 7. The results at 90°C are very similar and are given in the SI-1e. We observe that, mostly, the larger a pore, the more likely it is to reopen. We fitted a sigmoid function of the form $1/(1 + \exp(-(d - d_{char})/p))$ to this experimental data, where $d_{char}$ is a characteristic diameter and $p$ a characteristic standard deviation. A least-squares fit yielded $d_{char} = 33.6 \pm 3.5$ µm and $p = 4.0$ µm. Above $d_{char}$, the pores tend to reopen during drying, and below $d_{char}$ they tend to remain closed during drying. A fit of the same function to the data at 90 °C yielded $d_{char} = 26.0 \pm 2.0$ µm and $p = 5.1$ µm.

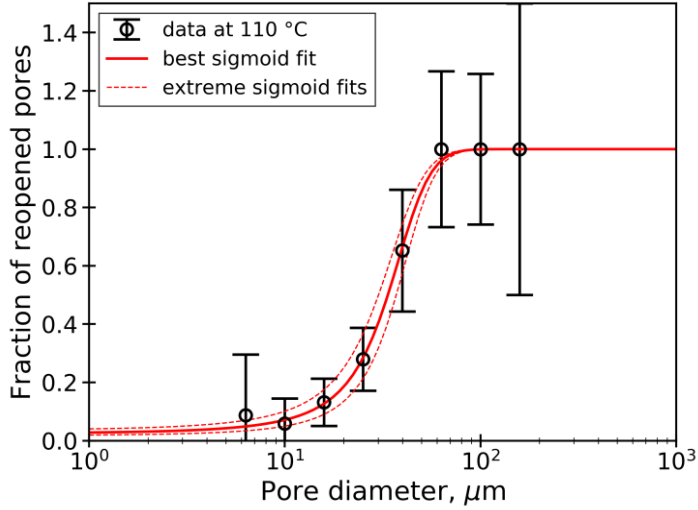

*Figure 7 – Fraction of pores in sample D200-5 that reopened after drying at 110°C. The data are obtained by adding results on 4 specimens. The parameters of the fitted sigmoidal function are $d_{char} = 33.6 \pm 3.5$ µm and $p = 4.0$ µm. The error bars are obtained as $1/\sqrt{N_i}$, where $N_i$ are the numbers of pores of diameter $d_i$ filled with water. The error on the fitted value of $d_{char}$ is estimated by finding out the extreme values of $d_{char}$ that make it still possible for the fitted sigmoid to remain in the experimental uncertainty.*

We now perform some hypothesis testing, by assuming that, in samples D300-30 and D70-30, the likelihood for a pore of a given diameter to reopen during drying at 110°C is the same as for sample D200-5 and is therefore given by the sigmoid function $F(d_i)$ displayed in Figure 7. We use the number-weighted distribution of pore diameters displayed in SI-1d, obtained from the cross-sections of the samples D300-30 and D70-30. Similarly to the derivation of the eq. 6, the probability of the pores to be found in the cross-section is inversely proportional to the pore diameter. Once all pores likely to reopen by cavitation have reopened, the number of reopened pores of diameter $d_i$ must be equal to $F(d_i)N_i^{section}$, so that we can estimate the porosity $\Phi_{reopen}$ that samples should have recovered during drying:

$$\Phi_{reopen} = \Phi_{initial} \frac{\sum_i F(d_i) N_i^{section} d_i^2}{\sum_i N_i^{section} d_i^2} \qquad (7)$$

The initial porosity of both samples D300-30 and D70-30 was $\Phi_{initial} = 30\%$. With Eq. (7), we estimate that, after drying at 110°C, the porosity of those 2 samples should be $\Phi_{reopen} = 26\%$ and $\Phi_{reopen} = 19\%$, respectively, while results displayed in Figure 6a show that the porosities of those 2 samples are only 3% and 17%, respectively. Therefore, we can conclude that the probability for pores to reopen after drying at 110°C is not the same for samples D300-30 and D70-30 as for sample D200-5. Consequently, this probability must not depend on the pore diameter only, but could depend also in particular on the porosity, since the initial porosity $\Phi_{initial}$ is the main difference between samples D300-30 and D200-5.

We now assume that, like for sample D200-5, there exists for samples D300-30 and D70-30 a characteristic size $d_{char}^{D300-30}$ and $d_{char}^{D70-30}$, respectively, above which pores tend to reopen during drying, and below which pores tend to remain closed during drying. Using Eq. (7) and the data for pore diameter distributions for samples D300-30 and D70-30, we find that correct predictions of the porosity after drying are obtained for $d_{char}^{D300-30} \sim 100 - 120$ μm and $d_{char}^{D70-30} \sim 100 - 120$ μm. This characteristic size is quite similar for samples D300-30 and D70-30 and therefore seems to not be impacted much by the mean diameter of the pores in the sample. This characteristic size is much larger than the one measured on sample D300-5, namely 34 μm. Consequently, we infer that the characteristic size (separating the larger pores that tend to reopen during drying from the smaller ones that tend to remain closed during this same drying) depends, for a given temperature of drying, on the porosity. Note that it could also depend on the mechanical properties of the solid skeleton or on the size of the sample (which impacts the kinetics of drying). The dependence of the characteristic size on the porosity may be due to the fact that the effective elastic modulus of the medium decreases with an increasing porosity. Indeed, we previously observed[9] a strong effect of porosity on the modulus of air-filled porous PDMS, i.e. a decrease by factor ~3 of the modulus of a sample with $\Phi = 30\%$ with respect to the non-porous one[*]. We thus may hypothesize that after reopening of one part of the porosity in the samples D300-30 and D70-30, pores that are still collapsed behave as if they are surrounded by an effective medium with much lower mechanical modulus, and hence the likelihood for those still-collapsed pores to reopen is significantly decreased. It is therefore possible that the characteristic size depends on the porosity as a consequence of its dependence on the elastic modulus of the medium.

Our observations do not allow to establish the mechanism of pore collapse and reopening. In accordance with the paper of Bruning et al.[12], we may expect that the condition necessary for cavitation in a water-filled pore should be that the (negative) pressure reaches a threshold value of about – 1.5 MPa. The existence of such threshold value, which is independent of the size of the pore, does not explain directly the pore size-dependence of the pore reopening that we observed. However, since cavitation of water is a probabilistic process, one could argue that cavitation is likelier to happen in larger pores (which contain a larger amount of water) than in smaller pores, which would be in agreement with our experimental observations. The fact that the likelihood of cavitation increases with temperature[13] would also be in agreement with our experimental observations. The second condition for pore reopening is that the tensile stresses that prevail at the location of the collapsed pore after cavitation of the fluid it contains has occurred should be high enough to allow the pore to reopen, akin to what is observed for "solid" cavitation (following the denomination that we used in the introduction). For small pores with $d < 0.1$ µm, the threshold tensile stress[16,17] for the growth of the pore may be much higher (in absolute value) than the Young's modulus $E = 1.6$ MPa (which provides an order of magnitude of the tensile stresses that must prevail at the location of the collapsed pore). If we assume that the cavitation takes place in water inside pores after they shrink to a diameter $d < 0.1$ µm, the surface energy may prevent the pore from the reopening after the cavitation occurred. This would explain the dependence of the cavitation on the pore diameter. However, it is difficult to directly correlate the size and shape of an initially spherical pore in a highly deformed state with its initial size.

---

[*] In the present manuscript, we measured a Young's modulus of wet samples with $\Phi = 30\%$ which is only slightly lower than that of wet samples with $\Phi = 5\%$ (0.98 MPa and 1.16 MPa, respectively), but such low effect of the porosity is due to the fact that, in our experiments, pores are filled with liquid water, whose bulk stiffness is very high (i.e. around 2 GPa).

*3.5. Drying in presence of NaCl*

In this section, we present microscopy observations during drying of samples which contain a NaCl salt solution inside pores. Due to its hydrophobic nature, the PDMS is impermeable to the salt[21]. Thus, drying leads to the concentration of the NaCl solution and crystallization of salt inside pores. The general image of drying is similar to the case without salt. However, a detailed investigation of the pore reopening shows the presence of cooperative effects.

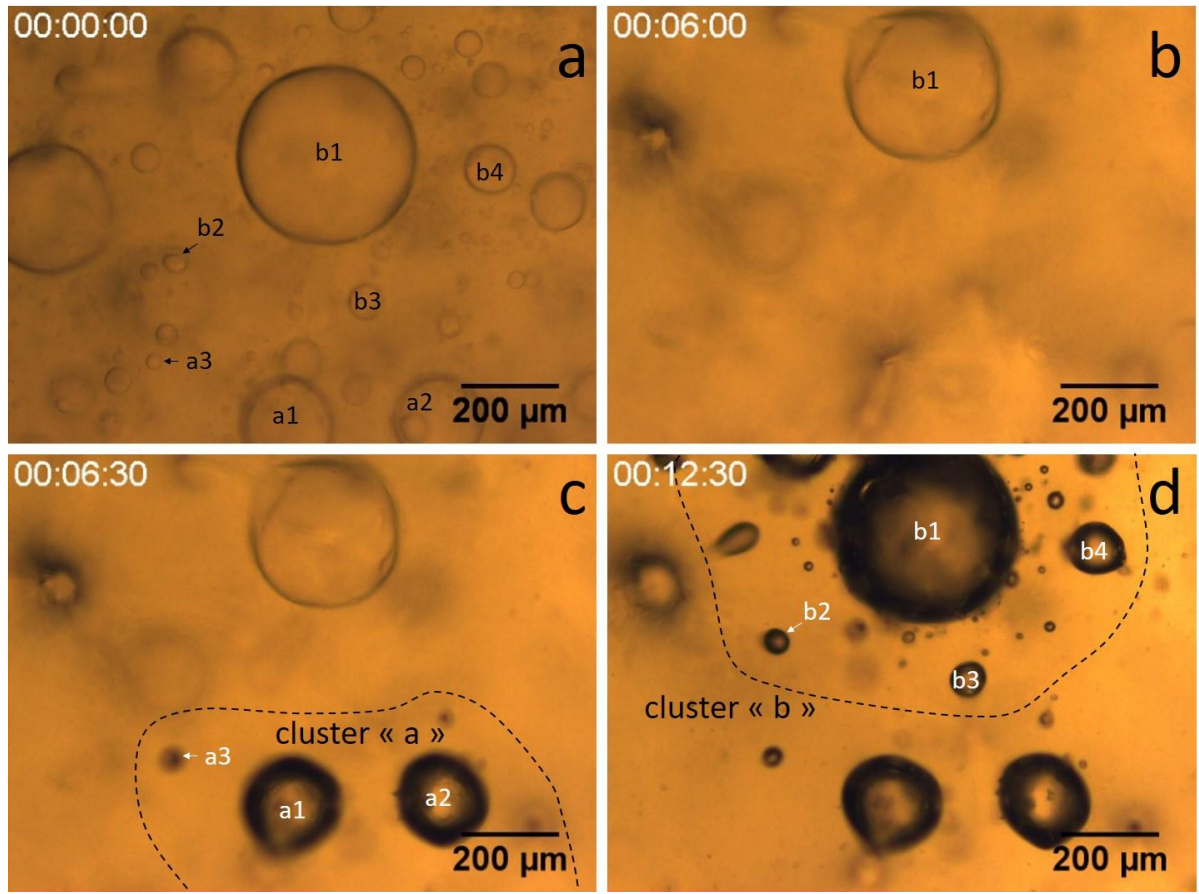

*Figure 8 – Optical microscopy images of different drying stages of porous PDMS sample with 5% porosity at 110°C. Symbols a1-a3 and b1-b4 mark similar pores in zones a and b, respectively, before and after cavitation.*

Figure 8 shows the optical images of the sample D300-5NaCl in which the pores are filled with a 1.5% wt. NaCl solution. For a video-stack of drying images, see the SI-5. The drying was performed at 110°C. At first, the pores shrink (Figure 8a and b), similarly to what was observed without salt (Figure 2c-f). No cavitation is observed during the first 6 minutes of drying. Next, a cluster of pores reopens simultaneously in the area labeled "a" (Figure 8c). In the following 6 minutes, no new cavitation events happen. Next, a new cluster reopens simultaneously in the area labeled "b" (Figure 8d). Similar qualitative features are observed for sample D50-5NaCl (whose pores are smaller than for sample D300-5NaCl and filled with 3.65% NaCl solution), for which results are presented on Figure 9. Again, the pore reopening happens by clusters of pores (see also SI-6 and SI-1f). Interestingly, in the center of clusters we observe the largest pores and the diameter of the cluster is about 2-3 times the diameter of these largest pores.

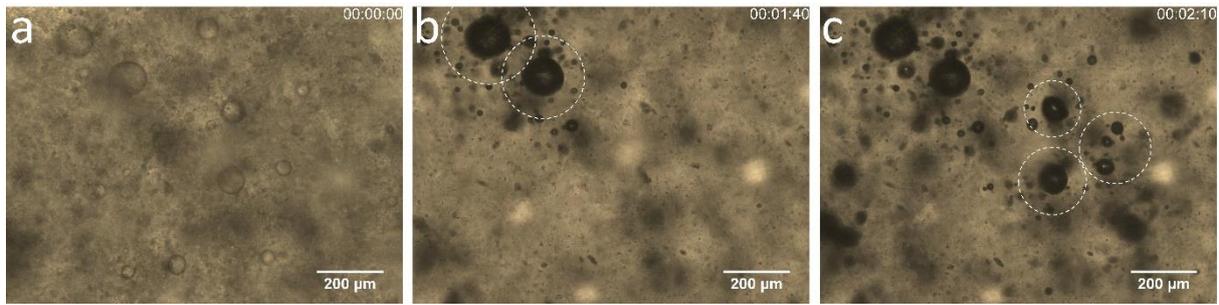

*Figure 9 – Optical microscopy images showing the reopening pore clusters during drying of the sample D50-5NaCl at 110°C.*

Figure 10a displays the diameter of some pores in the cluster "b" from Figure 8 versus the drying time. After creasing of the pore surface, we are not able anymore to measure a pore diameter; however, the times at which reopening occurs (which is indicated with star symbols on the figure) are clearly detected due to a good contrast of the vapor-filled pores. The dashed lines on Figure 10 are a linear extrapolation of the diameter evolution after creasing. While the collapse of small pores ("b2", "b3" and "b4") is expected to occur earlier than that of the large one ("b1"), we observe simultaneous reopening at about 900 s. Therefore, the reopening seems to be limited by the full shrinkage of the largest pore "b1". To better visualize this effect, we compare the reopening event statistics of samples D200-5 and D50-5NaCl on Figure 10b and 10c. Both samples contain statistically significant number of pores with diameters in the 10-100 μm range. The vertical bars correspond to the number of reopened pores between two imaging frames (about 10 s). Red bars show the events including at least one "large" pore with diameter > 50 μm. The comparison shows that in case of D200-5, many small pores reopen (black bars) independently of the large ones while in case of D50-5NaCl, most pores reopen simultaneously with the largest ones, i.e. are included in red bars. We attribute this difference to the osmotic pressure effect discussed in the following section.

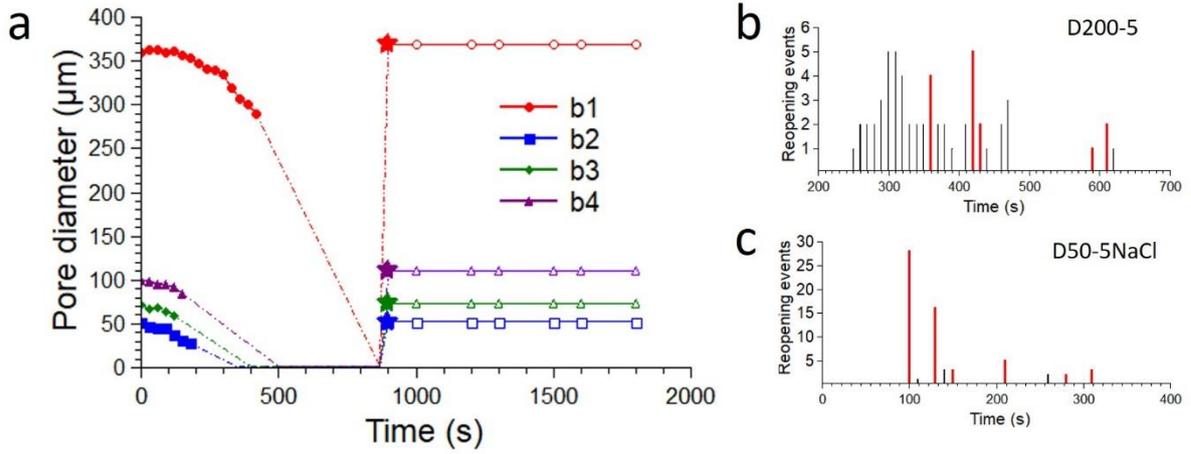

*Figure 10 – a) Evolution of the pore diameter before (filled symbols) and after cavitation (open symbols) for the pores from cluster "b" on Figure 8. The points represent the data obtained when the pores had a well-defined ellipsoidal shape and were located close to the focal plane of the microscope. The supposed evolution of pore diameters after creasing instability or after eventual displacement out of the focus are represented by dashed lines to guide the eye. The moments at which reopening occurs are indicated with the star symbol. b-c) Number of cavitation events as a function of drying time at 110°C for samples D200-5 (b) and D50-5NaCl (c). The red bars correspond to reopening events of at least one pore with diameter > 50 μm.*

*3.6. Discussion on the physical origin of the cooperative pore reopening*

In this section, we discuss why a cooperative pore reopening is observed in the samples whose pore solution contains NaCl. We will consider 2 factors successively: 1) osmotic pressure effects which slow the dissolution of small pores surrounding a large one; and 2) mechanical shocks caused by cavitation of a large pore that provoke the reopening of a cluster of pores.

We consider a large spherical pore "LP" of radius $R$ displayed in Figure 11a surrounded by a spherical shell of PDMS, which itself is surrounded by a "dry" air. The shell also contains a small pore "SP". In presence of salt, in any pore, the drying leads to an increase of the concentration $C$(NaCl) in salt in the pore which is inversely proportional to the volume $V(t)$ of this pore, as: $C(\text{NaCl}) = C_0(\text{NaCl}) \cdot V_0/V(t)$ where $C_0$(NaCl) and $V_0$ are the salt concentration and pore volume in the initial state.

As mentioned in section 3.2, in a diffusion-limited model of pore collapse, the time to full collapse of an individual pore scales with the square of the pore diameter. Hence, a smaller pore should collapse faster than a larger pore. However, this is only expected to be true if we neglect flow of water between neighboring pores: indeed, if a small pore is very close to a large pore, flow of water from the large pore to the small pore is expected to significantly slow down the collapse of the small pore. In contrast, if the small pore is closer to the outer edge of the shell, the impact of the large pore on the kinetics of collapse of the small pore is expected to be more limited. In the next paragraphs, we aim at estimating a characteristic length $L_{osm}$ of influence of the large pore. We define this length of influence as the distance to a large pore below which, because of the presence of the large pore, the salt concentration in the small pore cannot become larger than the solubility.

At first, we calculate the water concentration distribution that would prevail in the PDMS around the large pore if the shell around the large pore contained no small pore. The thickness of the shell is noted $L_{dry}$ and its diffusion coefficient (expressed in m$^2$.s$^{-1}$) is noted $D$. If we associate a spherical system of coordinates to the large pore with $r$ the radial coordinate, the concentration $c$ of water in the PDMS shell verifies: $c(R) = c_1$ and $c(R + L_{dry}) = c_2$. The concentrations $c_1$ and $c_2$ are equilibrium water concentrations in PDMS at water activities $a_w^{LP}$ and $a_w^{out}$, respectively, where $a_w^{LP}$ is the water activity inside the large pore "LP", and $a_w^{out}$ is the water activity outside. We assume ideality of the solution, such that $c_1 = k_D * a_w^{SP}$ and $c_2 = k_D * a_w^{out}$, where $k_D$ is a Henry's-type constant. For first-order estimations, literature data[19] show that this assumption is reasonable.

The flow of water through the PDMS shell verifies Fick's law $\underline{j} = -D\underline{\nabla}c$, where $\underline{j}$ (expressed in mol.m$^{-2}$.s$^{-1}$) is the diffusion flux. Combined with the mass balance, Fick's law yields the classical diffusion equation: $\partial c/\partial t = D\Delta c$. We estimate the flow by considering that a steady-state has been reached, from which follows that $\Delta c = 0$. Solving this equation with the above-mentioned boundary conditions yields:

$$c(r) = \frac{R \cdot (R + L_{dry})}{L_{dry}} \cdot \frac{c_1 - c_2}{r} + \frac{1}{L_{dry}} \left( c_2 \cdot (R + L_{dry}) - c_1 \cdot R \right) \quad (8)$$

As our drying was performed in dry air ($a_w^{out} \approx 0$), we may assume $c_2 \approx 0$, which gives:

$$c(r) = c_1 \cdot \frac{R}{L_{dry}} \cdot \left( \frac{R + L_{dry}}{r} - 1 \right) \quad (9)$$

Now, we consider that a small pore "SP" is located in the shell at the characteristic distance $L_{osm}$ from the surface of the large pore "LP", and that the salt concentration in this small pore is at solubility. The water activity in this small pore is noted $a_w^{SP}$. With the definition of the characteristic distance $L_{osm}$ that we introduced, the small pore must be in thermodynamic equilibrium with the surrounding PDMS.

We assume that the size of the small pore is negligible and that the water concentration distribution in the PDMS in the vicinity of the small pore is determined by the large pore and may be found using Eq. (9), for $r = R + L_{osm}$:

$$c(R + L_{osm}) = c_1 \cdot \frac{R}{L_{dry}} \cdot \left( \frac{L_{dry} - L_{osm}}{R + L_{osm}} \right) \quad (10)$$

After taking $c(R + L_{osm}) = k_D * a_w^{SP}$ and $c_1 = k_D * a_w^{LP}$, we obtain the equation for $L_{osm}$:

$$L_{osm} = \left( \frac{R}{a_w^{SP} + a_w^{LP} R/L_{dry}} \right) (a_w^{LP} - a_w^{SP}) \quad (11)$$

Equation 11 may be simplified if $R \ll L_{dry}$, which is a strong but reasonable assumption for pores with $R \sim 100$ μm observed in depth of a rectangular sample with dimensions 10 mm × 10 mm × 0.5 mm (i.e. $L_{dry} \sim 250$ μm):

$$L_{osm} \approx R \left( \frac{a_w^{LP}}{a_w^{SP}} - 1 \right) \quad (11a)$$

We consider that the NaCl concentration in the small pore reached its solubility: its concentration $C_{wt}$ in g per g of solution is 27 wt%. The NaCl concentration in the large pore is taken to be the one prevailing in sample D300-5NaCl initially, which is equal to 1.5 wt%. Since, assuming ideality, the water activity $a_w$ verifies $a_w = (1 - C_{wt})/(1 - C_{wt} + 2 * C_{wt} * M_{H_2O}/M_{NaCl})$, where $M_{H_2O} = 18.0$ g.mol$^{-1}$ and $M_{NaCl} = 58.4$ g.mol$^{-1}$ are the molar mass of water and NaCl, respectively, the water activities corresponding to those concentrations are $a_w^{SP} = 0.81$ and $a_w^{LP} = 0.99$. The application of equation 11 (for $R = 100$ μm and $L_{dry} = 250$ μm) and its approximate version 11a give $L_{osm} = 14.5$ μm and $L_{osm} = 21.6$ μm, respectively. If the small pore "SP" is at a distance larger than $L_{osm}$ from the large pore "LP", it can further shrink. But, in contrast, if it is at a distance smaller than $L_{osm}$ from the large pore "LP", the small pore "SP" will in rather gain water and will not go on shrinking until the large pore has significantly shrunk.

In presence of multiple large pores, as on Figure 9, the dissolution of small pores between those large pores will be strongly slowed down until the large pore reaches the solubility limit of NaCl, at which point the water in large and small pores will be roughly the same and all osmotic effects will have vanished. Until then, osmotic effects may keep the magnitude of the negative pressure in the small pores low enough (in absolute value) to prevent cavitation. Such mechanism can explain that, in presence of NaCl, very few small pores reopen before the neighboring large pore has shrunk and is prone to cavitation itself (see Figure 10c). In absence of NaCl, there is no osmotic effect between neighboring pores, such that small pores are less impacted by neighboring large pores and can cavitate before the neighboring large pore has itself shrunk (see Figure 10b).

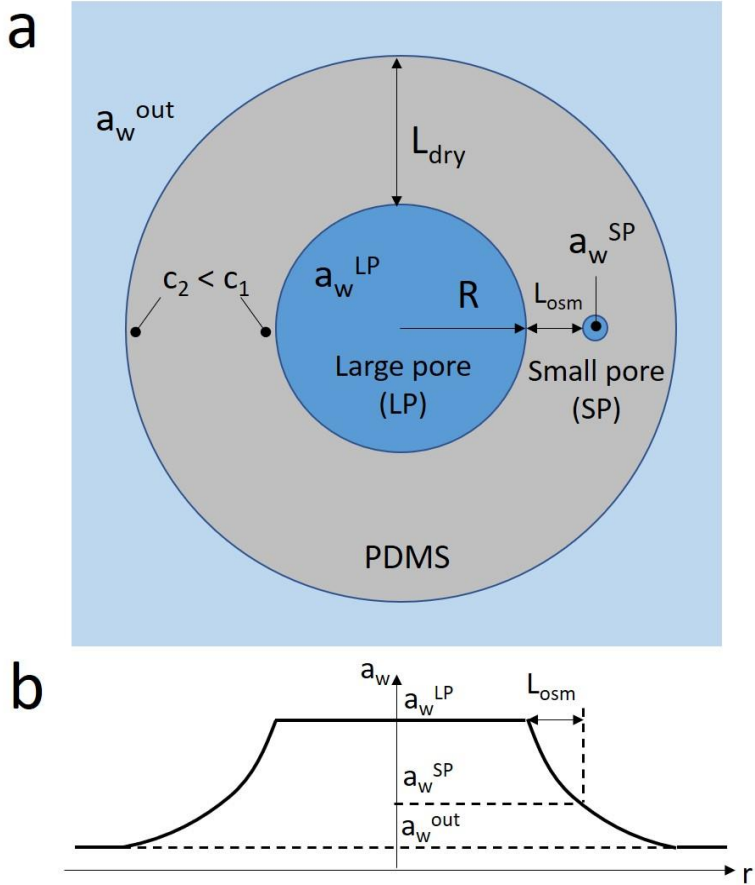

*Figure 11 –a) The model used to estimate the osmotic length $L_{osm}$: a small spherical pore "SP" with high concentration of NaCl is located near a large spherical pore "LP" in a PDMS shell of thickness $L_{dry}$. $a_w^{LP}$ and $a_w^{SP}$ and $a_w^{out}$ are the activities of water in the large pore, small pore and outside medium, respectively; $c_1$ and $c_2$ design the water concentrations in PDMS near the large pore and near the border. The schematic is not drawn to scale, in the sense that the characteristic distance $L_{dry}$ of the large pore to the edge of the sample is, for most pores, much larger than the size of the pore b) The corresponding schematic of the radial distribution of water activity in the shell, when neglecting the impact of the small pore "SP".*

The second factor explaining the observation in section 3.5 of a simultaneous reopening of neighboring pores is based on the recent work of Doinikov et al.[22]. These authors proposed a model which predicts that the cavitation in a fluid-filled pore under negative pressure may provoke a cavitation in the neighboring pore at the distance of the order of the pore radius. This effect is due to the mechanical shock produced by the cavitation, which may instantly increase the absolute value of the negative pressure inside the neighboring pore. As shown by Vincent et al.[13] for rigid water-filled pores, the cavitation is very sensitive to the pressure and even relatively small changes may strongly increase the cavitation probability.

In conclusion, we explain the observed cooperative pore reopening in presence of NaCl by two effects: the osmotic pressure which tends to equilibrate the negative pressure between the neighboring large and small pores; and the mechanical shock transmission from a pore in which cavitation occurs to its neighbors.

## 4. Conclusions

In this paper, we investigated the drying of water-filled PDMS sponges with 5% and 30% of closed porosity. In these materials, the drying may lead either to a collapsed state with low porosity or to the cavitation and reopening of a fraction of the pores. Using optical microscopy and porosity measurements, we showed the importance of the pore diameters and interactions on the result of drying. At pore diameters lower than 20 μm, the majority of pores remain collapsed. This is most likely due to the adhesion of the pore walls and low probability of the cavitation in a pore with a small volume. Pores with diameters larger than 100 μm tend to reopen after the fluid they contain cavitates. The behavior of pores with diameters ranging from 20 to 100 μm depends on the porosity and drying temperature. In the last sections, we showed evidence of cooperative pore reopening in samples containing NaCl solution inside their pores. We explained this effect by the action of osmotic pressure and the transmission of mechanical shock from a pore in which cavitation occurs to its neighbors.

Our results show the complexity of the drying process for highly deformable materials such as elastomer sponges. Many open questions remain including whether small amounts of water are still present inside the collapsed pores or not, or what the role of adhesion and mechanical damage of the elastomer matrix on the collapse/reopening process is. To better understand the influence of pore sizes, one needs experiments on monodisperse sponges[23,24] with controllable diameter distributions and porosities. We did not explore porosities larger than 30%, for which we may expect stronger cooperative effects between pores and the existence of collective buckling modes similar to those observed for polymer foams[25].

## 5. Author Information

Corresponding author:

* E-mail: artem.kovalenko@espci.fr

Notes : The authors declare no competing financial interest.

## 6. Acknowledgments

We thank Andrea Aguiar for her help in formulation of the samples, Ludovic Olanier for help with the fabrication of the hydrostatic balance and Etienne Barthel for fruitful discussions. We also thank the DIM Respore ( http://www.respore.fr/ ) for having given the last 2 authors the opportunity to meet.

**References:**

(1) Zhu, D.; Handschuh-Wang, S.; Zhou, X. Recent Progress in Fabrication and Application of Polydimethylsiloxane Sponges. *J. Mater. Chem. A* **2017**, *5* (32), 16467–16497. https://doi.org/10.1039/C7TA04577H.

(2) Brunet, T.; Merlin, A.; Mascaro, B.; Zimny, K.; Leng, J.; Poncelet, O.; Aristégui, C.; Mondain-Monval, O. Soft 3D Acoustic Metamaterial with Negative Index. *Nature Materials* **2015**, *14* (4), 384–388. https://doi.org/10.1038/nmat4164.

(3) Jin, Y.; Kumar, R.; Poncelet, O.; Mondain-Monval, O.; Brunet, T. Flat Acoustics with Soft Gradient-Index Metasurfaces. *Nature Communications* **2019**, *10* (1), 143. https://doi.org/10.1038/s41467-018-07990-5.

(4) Kovalenko, A.; Fauquignon, M.; Brunet, T.; Mondain-Monval, O. Tuning the Sound Speed in Macroporous Polymers with a Hard or Soft Matrix. *Soft Matter* **2017**, *13* (25), 4526–4532. https://doi.org/10.1039/C7SM00744B.

(5) Silverstein, M. S. PolyHIPEs: Recent Advances in Emulsion-Templated Porous Polymers. *Progress in Polymer Science* **2014**, *39* (1), 199–234. https://doi.org/10.1016/j.progpolymsci.2013.07.003.

(6) Tebboth, M.; Jiang, Q.; Kogelbauer, A.; Bismarck, A. Inflatable Elastomeric Macroporous Polymers Synthesized from Medium Internal Phase Emulsion Templates. *ACS Appl. Mater. Interfaces* **2015**, *7* (34), 19243–19250. https://doi.org/10.1021/acsami.5b05123.

(7) Turani-i-Belloto, A.; Meunier, N.; Lopez, P.; Leng, J. Diffusion-Limited Dissolution of Calcium Carbonate in a Hydrogel. *Soft Matter* **2019**, *15* (14), 2942–2949. https://doi.org/10.1039/C8SM02625D.

(8) Martina, A. D.; Hilborn, J. G.; Kiefer, J.; Hedrick, J. L.; Srinivasan, S.; Miller, R. D. Siloxane Elastomer Foams. In *Polymeric Foams*; ACS Symposium Series; American Chemical Society, 1997; Vol. 669, pp 8–25. https://doi.org/10.1021/bk-1997-0669.ch002.

(9) Kovalenko, A.; Zimny, K.; Mascaro, B.; Brunet, T.; Mondain-Monval, O. Tailoring of the Porous Structure of Soft Emulsion-Templated Polymer Materials. *Soft Matter* **2016**, *12* (23), 5154–5163. https://doi.org/10.1039/C6SM00461J.

(10) Milner, M. P.; Jin, L.; Hutchens, S. B. Creasing in Evaporation-Driven Cavity Collapse. *Soft Matter* **2017**, *13* (38), 6894–6904. https://doi.org/10.1039/C7SM01258F.

(11) Cai, S.; Bertoldi, K.; Wang, H.; Suo, Z. Osmotic Collapse of a Void in an Elastomer: Breathing, Buckling and Creasing. *Soft Matter* **2010**, *6* (22), 5770. https://doi.org/10.1039/c0sm00451k.

(12) Bruning, M. A.; Costalonga, M.; Snoeijer, J. H.; Marin, A. Turning Drops into Bubbles: Cavitation by Vapor Diffusion through Elastic Networks. *Phys. Rev. Lett.* **2019**, *123* (21), 214501. https://doi.org/10.1103/PhysRevLett.123.214501.

(13) Vincent, O.; Sessoms, D. A.; Huber, E. J.; Guioth, J.; Stroock, A. D. Drying by Cavitation and Poroelastic Relaxations in Porous Media with Macroscopic Pores Connected by Nanoscale Throats. *Physical Review Letters* **2014**, *113* (13), 134501. https://doi.org/10.1103/PhysRevLett.113.134501.

(14) Wheeler, T. D.; Stroock, A. D. The Transpiration of Water at Negative Pressures in a Synthetic Tree. *Nature* **2008**, *455* (7210), 208–212. https://doi.org/10.1038/nature07226.

(15) Fond, C. Cavitation Criterion for Rubber Materials: A Review of Void-Growth Models. *Journal of Polymer Science Part B: Polymer Physics* **2001**, *39* (17), 2081–2096. https://doi.org/10.1002/polb.1183.

(16) Gent, A. N.; Tompkins, D. A. Surface Energy Effects for Small Holes or Particles in Elastomers. *Journal of Polymer Science Part A-2: Polymer Physics* **1969**, *7* (9), 1483–1487. https://doi.org/10.1002/pol.1969.160070904.

(17) Dollhofer, J.; Chiche, A.; Muralidharan, V.; Creton, C.; Hui, C. Y. Surface Energy Effects for Cavity Growth and Nucleation in an Incompressible Neo-Hookean Material—

Modeling and Experiment. *International Journal of Solids and Structures* **2004**, *41* (22), 6111–6127. https://doi.org/10.1016/j.ijsolstr.2004.04.041.

(18) Stricher, A. M.; Rinaldi, R. G.; Barrès, C.; Ganachaud, F.; Chazeau, L. How I Met Your Elastomers: From Network Topology to Mechanical Behaviours of Conventional Silicone Materials. *RSC Adv.* **2015**, *5* (66), 53713–53725. https://doi.org/10.1039/C5RA06965C.

(19) Harley, S. J.; Glascoe, E. A.; Maxwell, R. S. Thermodynamic Study on Dynamic Water Vapor Sorption in Sylgard-184. *J. Phys. Chem. B* **2012**, *116* (48), 14183–14190. https://doi.org/10.1021/jp305997f.

(20) Cheng, H. C.; Lemlich, R. Errors in the Measurement of Bubble Size Distribution in Foam. *Ind. Eng. Chem. Fundamen.* **1983**, *22* (1), 105–109. https://doi.org/DOI: 10.1021/i100009a018.

(21) Naeimi, M.; Karkhaneh, A.; Barzin, J.; Khorasani, M. T.; Ghaffarieh, A. Novel PDMS-Based Membranes: Sodium Chloride and Glucose Permeability. *Journal of Applied Polymer Science* **2013**, *127* (5), 3940–3947. https://doi.org/10.1002/app.37709.

(22) Doinikov, A. A.; Dollet, B.; Marmottant, P. Cavitation in a Liquid-Filled Cavity Surrounded by an Elastic Medium: Intercoupling of Cavitation Events in Neighboring Cavities. *Phys. Rev. E* **2018**, *98* (1), 013108. https://doi.org/10.1103/PhysRevE.98.013108.

(23) Giustiniani, A.; Guégan, P.; Marchand, M.; Poulard, C.; Drenckhan, W. Generation of Silicone Poly-HIPEs with Controlled Pore Sizes via Reactive Emulsion Stabilization. *Macromolecular Rapid Communications* **2016**, *37* (18), 1527–1532. https://doi.org/10.1002/marc.201600281.

(24) Wang, B.; Prinsen, P.; Wang, H.; Bai, Z.; Wang, H.; Luque, R.; Xuan, J. Macroporous Materials: Microfluidic Fabrication, Functionalization and Applications. *Chem. Soc. Rev.* **2017**, *46* (3), 855–914. https://doi.org/10.1039/C5CS00065C.

(25) Gibson, L. J.; Ashby, M. F. The Mechanics of Three-Dimensional Cellular Materials. *Proceedings of the Royal Society A: Mathematical, Physical and Engineering Sciences* **1982**, *382* (1782), 43–59. https://doi.org/10.1098/rspa.1982.0088.